# MS-Twins：Multi-Scale Deep Self-Attention Networks for Medical Image Segmentation

**Jing Xu[1], ***

1. Department of Computing, Imperial College London, London, UK
* j.xu23@imperial.ac.uk

**Abstract**: Although transformer is preferred in natural language processing, some studies has only been applied to the field of medical imaging in recent years. For its long-term dependency, the transformer is expected to contribute to unconventional convolution neural net conquer their inherent spatial induction bias. The lately suggested transformer-based segmentation method only uses the transformer as an auxiliary module to help encode the global context into a convolutional representation. How to optimally integrate self-attention with convolution has not been investigated in depth. To solve the problem, this paper proposes MS-Twins (Multi-Scale Twins), which is a powerful segmentation model on account of the bond of self-attention and convolution. MS-Twins can better capture semantic and fine-grained information by combining different scales and cascading features. Compared with the existing network structure, MS-Twins has made progress on the previous method based on the transformer of two in common use data sets, Synapse and ACDC. In particular, the performance of MS-Twins on Synapse is 8% higher than SwinUNet. Even compared with nnUNet, the best entirely convoluted medical image segmentation network, the performance of MS-Twins on Synapse and ACDC still has a bit advantage.

## 1 Introduction

Convolution neural network (CNN), as a core technology of many computer vision systems, has significantly advanced the state-of-the-art of image analysis in the past decade. Recently, Transformers [1], which originally deals with natural language processing, has received extensive attention in vision-based research and applications [2-3]. The central concept behind the transformer is to use the self-attention mechanism to learn long-range dependencies in data representations. In comparison to CNN, the transformer is not confined to the local induction bias enforced by the convolutional kernel, which makes it more suitable for learning nonlocal global context [5]. However, when the transformer is applied to vision tasks, it needs to take image patches as the basic units in representations and thus loses detailed pixel-level information [Cite Vision Transformer], which CNN is better at. It is found that the prediction error of Transformers is closer to the forecast error of human beings than that of CNN.

Given the respective advantages of transformers and CNN, numerous works have been proposed to develop a hybrid architecture that integrates transformers with CNN for medical image segmentation. Chen et al. [6] proposed TransUNet for the first time to explore the potential of transformers in medical image segmentation. TransUNet follows a U-shaped architecture design, similar to that of U-Net [7]. It employs the transformer to encode global context in the encoder path, which is then upsampled and integrated with fine-detail CNN features in the decoder path. TransUNet and most of its followers [8-11] only regard the convolution neural network as the subject, and further apply Transformers on the subject to capture the long-term dependency. Because the convolution representation usually contains accurate spatial information and provides hierarchical concepts, one or two layers of Transformer are not enough to combine the long-term dependency with the convolution representation, so the advantages of Transformer are not brought into full play. Some works [12-14] proposed to use the transformer for both the encoder and decoder paths of the segmentation model. Inspired by the Swin Transformer [3], Swin-UNET [12] uses layered transformer blocks to build encoders and decoders in an architecture similar to U-Net, which demonstrates an improved performance compared to TransUNet. However, it does not probe proper bond convolution and self-attention to build a top medicine partition network.

Recently, Zhou et al. [16] proposed a volume segmentation model, nnFormer (bonds convolution and self-attention for the first time), which is not used in the traditional voxel-based self-attention computing model, but adopts a local 3D image block calculation method, and its backbone is based on Swin Transformer proposed by Liu [3]. Swin Transformer is a local attention-based model with low computational complexity, but it also loses the global sensory field modeling ability of global attention. Twins [16], newly proposed by Chu, combines local attention and global attention and improves the global modeling ability based on a single local attention model by alternately using LSA (local group self-attention) and GSA (global sub-sample attention).

Therefore, this paper proposed a multi-scale deep self-attention segmentation network. Based on the Twins Transformer block and multi-scale feature iterative fusion block, iteratively fuse features of different scales. Construct a cascading feature extraction structure, predicting objects that were not correctly predicted at the previous level from top to bottom layer by layer. A large number of tests on multiple and heart partition data sets show that this way has good segmentation precision. Specifically, the contributions can be concluded as follows: (1) Proposed a powerful segmentation model-MS-Twins, on account of the Twins Transformer block and the previously proposed multi-scale features iterative fusion MS-FIF (Multiscale Feature Iterative Fusion) module, through this iterative fusion of different scale features to obtain more meaningful information. (2) Construct a cascaded feature extraction structure to predict the incorrectly predicted objects in the upper level from high to low, to obtain more precise targets. (3) From the point of view of vision perceive and data distribution, a novel loss function is put forward, which consists of a contrastive loss and a balance loss. The first mentioned of two can alleviate the difference between different categories of each sample. The latter can help the model learn to predict the final goal more accurately.

In the experimental part, this study compares MS-Twins with various baseline segmentation methods. In the multi-organ segmentation task of Synapse, the proposed MS-Twins algorithm is more than 8% higher than the SwinUNet algorithm. On average, MS-Twins performs nearly 1% better than SwinUNet when performing automatic heart diagnostics on ACDC data sets. It can be seen that MS-Twins has achieved good results in the research of medical image segmentation.

## 2 Related work

In the section, the work review recently proposed methods that use transformers to perform medical image segmentation. Most of them use a mixed architecture of convolution and transformer [1]. This study group them into two types according to if most of the backbone is convolution-based or transformer-based.

**Convolution-based backbone.** TransUNet [6] follows a U-net style architecture. It first performs convolution to generate feature maps for image patches, which are tokenised and fed into transformer layers to extract global context. The global context is then upsampled and combined with CNN feature maps in similar way to the U-net decoder. At a similar time to TransUNet, Li et al. [17] proposes to apply transformers to CNN feature maps to learn global context. In particular, a squeeze-and-expansion transformer is proposed, in which the squeezed attention block is designed to be more compact than the vanilla attention block and the expanded attention block aims to increases model capacity to learn more diversified representation of input images. TransFuse [11] proposes the BiFusion block to combine the features from the CNN encoder and the transformer encoder, followed by prediction of the 2D segmentation map. Compared with TransUNet, TransFuse almostly applies the self-attention mechanism to the import implanting tier to promote the 2D image segmentation model. Yun et al. [18] employs the transformer to learn representation for both spatial and spectral information, which is designed specifically to model the context in the spectral bands in hyperspectral pathology images. Xu et al. [19] introduced an efficient encoder, LeViT module, into a U-net architecture with the aim to achieve a trade-off between accuracy and efficiency for the segmentation model. Li et al. [20] proposed a novel upsampling method, window attention upsample, which is implemented in the decoder part of the transformer. TransClawU-network [8] combines convolution with transformer in the encoder path of a U-net. TransAttUNet [9] developed a novel self-aware attention module with transformer self-attention and global spatial attention, which is integrated into a U-net architecture (GSA). CoTr [21] used a deformable transformer to process multi-scale features extracted from convolutional layers at different layers. TransBTS [22] first uses 3D-CNN to extract volumetric feature maps, which are downsampled to form the input tokens for the transformer to model global feature relationships. Different from the methods which directly use convolution for extracting features which are then fed to the transformer, the proposed MS-Twins interlaces the convolution and transformer blocks, so that they make use of each other in feature extraction.

**Transformer-based backbone.** Valanarasu et al. [10] proposed the medical transformer (MedT) which uses a gated axial attention layer as the basic building blocks and constructs two branches for feature learning, namely a global branch and a local branch. Karimi et al. [13] proposed a convolution-free architecture which only uses transformer blocks for 3D medical image segmentation. Swin-UNET [12] proposed a U-shaped encoder-decoder architecture, where the basic building block is the Swin transformer [3], which uses multi-scale shifted windows to extract hierarchical features. DS-TransUNet [14] further extends Swin-UNET, which uses two Swin transformer encoder branches to learn features for image patches at two different scales, then fuses their features and feeds to the decoder branch. nnFormer [15] (not-another transFormer) proposed a computationally efficient 3D transformer framework for volumetric medical image segmentation which combines convolution and self-attention operations in feature extraction. MS-Twins obtains the advantages of convolution in coding accurate spatial data and generating graduated representation that helps to model object conception at different dimensions and better integrates self-attention and convolution neural networks.

The other parts of the paper are described as follows. In this third part, the proposed techniques are introduced in detail, including deep segmentation network, multi-scale feature iterative fusion module, and multi-scale prediction Loss function. In the fourth section, the experimental results and analysis are given. Finally, the fifth section is the conclusion.

## 3 Method

### 3.1 Overview

The overall framework of MS-Twins is shown in Figure 1(A), which is composed of two branches, namely, encoder and decoder. The encoder includes four Twins transformer modules; symmetrically, the decoder branch also includes four Twins transformer modules. Inspired by U-Net [7] and FractalNet [23], this study symmetrically adds the multi-scale feature iterative fusion module between the feature pyramids corresponding to the encoder and decoder, which is helpful to iteratively fuse the feature result graphs extracted by different layers. The prediction output of each layer of the characteristic pyramid corresponding to the decoder is compared with the remaining incorrectly predicted samples in the previous layer, the residual is uploaded step by step, and the fine-grained details in the prediction are restored.

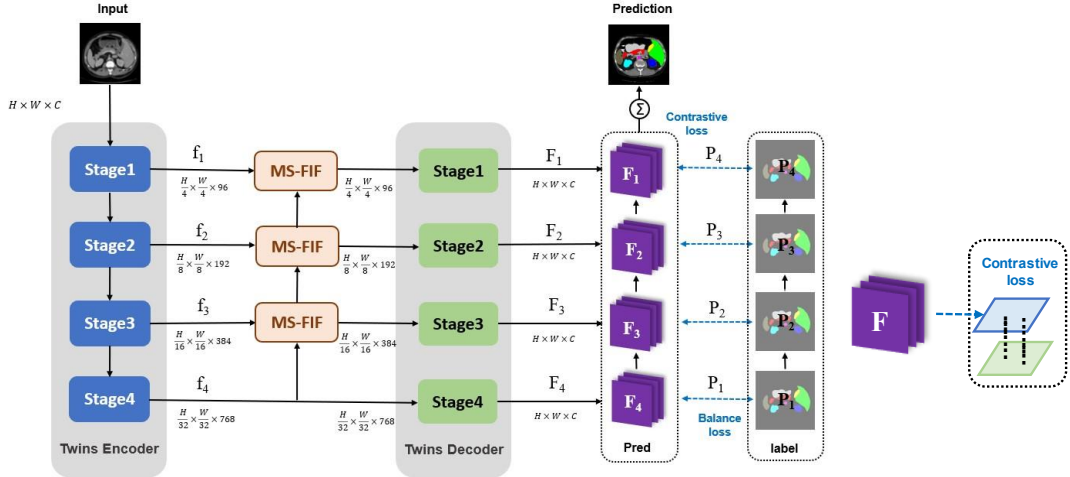

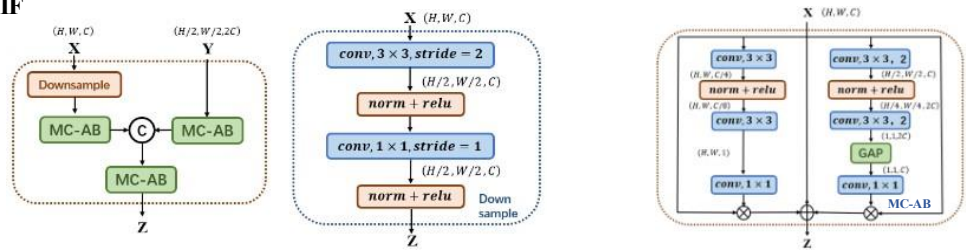

Fig. 1: Architecture of MS-Twins. (A) The whole framework of MS-Twins is shown. (B) More details about MS-FIF are given. Notice that the architecture shown applies to the images from the pre-processed Synapse dataset. Depending on the size of the input patch, the architecture may be slightly different.

### 3.2 Encoder

In the encoder, when the data passes through four Twins transformer modules of the stage, the size of each feature will be halved, and the number of channels of the feature will be increased by two times. Taking the second stage as an example, the input feature $\left(\frac{H}{4} \times \frac{W}{4} \times 96\right)$ is changed to

$\left(\frac{H}{8} \times \frac{W}{8} \times 192\right)$ after patchEmbed, the input twins-transformer module is used for feature extraction, and the output $\left(\frac{H}{8} \times \frac{W}{8} \times 192\right)$ is output to the next stage and MS-FIF.

### 3.3 MS-FIF

The various channel focus module MC-AB (Multi-Channel Attention Block), as demonstrated in Figure 1(B), possesses a relatively not complicated construction and applies two parts with distinct scales to fetch channel focus weights. One part fetches the spatial focus of the local feature, and the other applies Universal Avg Pooling averaging to fetch the channel focus of the universal feature.

As demonstrated in Figure 1(B) above-mentioned, the multi-scale feature iterative fusion module MS-FIF (Multiscale Feature Iterative Fusion) mainly aims at the attention problem of various scale feature fusion in distinct network frameworks. Shown two characteristic graphs $X$, $Y \in R$, $Y$ are characteristic graphs with a wide range of receptive fields, where MC-AB is a various channel focus module, $Z \in R^{C \times H \times W}$ is the export character after fusion, and $C$ represents combination. The massive character images are down sampled so that these two-character images are same scale, and the recovered ones are connected through a various channel focus module respectively, and finally export through a various channel focus module.

The features of the two-stage adjacent to the encoder are input through MS-FIF, for example, the fainter 3 of the stage3 output $\frac{H}{16} \times \frac{W}{16} \times 384$ and the stage4 output $\frac{H}{32} \times \frac{W}{32} \times 768$ are inputted into the MS-FIF, and the feature result of the fused $\frac{H}{16} \times \frac{W}{16} \times 384$ is obtained, and then the corresponding level of the decoder is input, and the other stage is the same.

### 3.4 Decoder

The structure of the Transformer block of the decoder is almost symmetrical with that of this encoder. After MS-FIF fusion, the feature layer is fed into the corresponding stage module and tagged and predicted. Because our method is to achieve multi-level approximation prediction, high-level features have more abstract features, so this study use high-level features to do primary target prediction, and the remaining incorrect targets are predicted through low-level feature decoders. Because the low-level features are fine-grained features, they are used to predict the incorrectly predicted objects at the upper level, to achieve the direction of multi-level fine entropy reduction, to better capture semantic and fine-grained information.

### 3.5 Loss design

**Contrastive Loss**. There are many kinds of problems in data segmentation regions, which are not good to the network to distinguish heterogeneous texture data; most samples cover diverse types of segmentation regions at the same time, which is hard to drill; the segmentation data of various regions is distinct, which leads to various levels of learning. This study has designed a loss function for comparison, inspired by YannLeCun et al. [24]. Through this loss, MS-Twins model can not only increase the differences between classifiers, but also mine more meaningful features from the relationship of paired data. This study summarizes the contrast loss of each category region in the data to represent the contrast loss of multi-segmented region data.

$$L_{con} = \sum_{i}^{n} \min\left(\frac{2|X \cap Y|}{|X|+|Y|}\right) \quad (1)$$

Among them, $X$, $Y$ is the predicted value of different categories after the fusion of each level, and $i$ represents the level. Where $|X \cap Y|$ is the intersection of $X$ and $Y$, and $|X|$ and $|Y|$ can be added directly or summed by the square of elements respectively. The coefficient of the numerator is 2 because there is a repeated calculation of the intersection between $X$ and $Y$ by the denominator.
**Balance Loss**. To further mine the more meaningful features in the multi-scale model, this study defines a balance loss. In figure 1 (A), this study first obtains the last layer result of the decoder output on the high-level feature, balance the loss with the mark feature, and calculate the balance loss between the sub-high-level feature and the prediction result output and the mark of the upper-level incorrectly predicted object. Similarly, the high-level features are calculated step by step to the lower-level features, to achieve the direction of multi-level fine entropy reduction. The formula is as follows:

$$L_{bal} = \sum_{j=1}^{4} \sum_{i=1}^{c} (1 - \widehat{P_i})^{q_i} \log(\widehat{P_i}) \quad (2)$$

Where $c$ is the total quantity of kinds and $i$ represent each label of different segmented regions. $\widehat{P_i}$ ($i = 1,2,\cdots,14$) show the probability value (predicted value) of the layer $i$ feature predicted by the network. Parameter $q_i$ is a hard and easy sample element, which may have better information of mining features under some $q_i$ settings, thus improving model drilling.
**Multi-Scale Prediction Loss**. The various scale prediction loss function can be a combined loss function, that is:

$$L = \alpha L_{con} + L_{bal} \quad (3)$$

Where $\alpha$ represents the tradeoff coefficient between the contrastive loss and the balance loss. contrastive loss will prompt the model to pay more accurate attention to the segmentation of different classification areas. The balance loss helps the model to learn to predict the target better.

## 4 Experiments

In order to compare MS-Twins fairly with former Transformer-based frameworks, this study conducted tests on multiple organ CT segmentation challenges (Synapse) [25] and automatic heart diagnosis challenges (ACDC) [26] datasets.
**Synapse for multi-organ CT segmentation.** The dataset included 3,779 abdominal axial clinical CT images of 30 subjects. Following the practice in [27], the dataset was split into 18 subjects for training and 12 subjects for testing. Eight abdominal organs (spleen, aorta, left kidney, right kidney, gallbladder, pancreas, stomach and liver) were evaluated using the Dice similarity coefficient (Dice).
**ACDC for automated cardiac segmentation.** ACDC involved 150 sick persons The left ventricular cavity (LV), left ventricular myocardium (Myo) and right ventricular cavity (RV) were manually segmented. The dataset was split into 70 training samples, 10 validation samples and 20 test samples. Again, use the average Dice to evaluate the approach.

### 4.1 Implementation details

All the experiments were based on Python3.6, PyTorch 1.7.1, and Ubuntu 18.04. The whole testing steps are acted on an NVIDIA 3090 GPU using 24 GB memory. The primary learning speed is put to 0. 01, the default optimize procedure is SGD, and put the momentum to 0. 9. Weight fall off is made to 1e-4.

**Pre-processing and data augmentation.** The whole images are first resampled to the identical target spacing. In the preprocessing process, enhancements such as brightness and contrast adjustment, rotation, low-resolution simulation, scaling, gamma enhancement, mirroring, Gaussian noise, and Gaussian blur are applied in a given order.

**Pre-trained model weights.** Pre-training is vital to apply a general and translatable representation for downstream assignments. Considering that many operations in MS-Twins handle on one-dimensional lists, this study find out the possibility of transferring the pre-trainied weights from natural graphics to the field of medicine imaging. Even more specifically, this study's goal is to gain the weight of the pre-trained MLP layer in ADE 20K pre-training. To do this, this study align the number of channels of the Transformers block with the number of channels of the pre-trained model to load the weight of the MLP layer.

**4.2 Experiments on Synapse**

As listed in Table 1, Experiments about synapse were conducted and compared the MS-Twins with various transformer and convnet on account of baselines. The main estimate index is the dice mark.

In addition to nnUNet, the convnet on account of way of the best performance is DualNormUNet [28], with a mean dice mark of 80.37%. By comparison, the transformer-based consequnces reported by WAD can be the best, with an average of 80.30%, slightly lower than DualNormUNet. The MS-Twins average score is 7.79% and 7.72% higher than WAD and DualNormUNet respectively and achieved SOTA effect on synapse data. In addition, this study finds that the performance of nnUNet is seriously underestimated. After careful adjustment, nnUNet's mean dice mark reached 86.99%, better than DualNormUNet and WAD, but still inferior to the suggested MS-Twins.

Table 1: Experiments on Synapse (dice mark in %, Best consequences are bolded.)

| Methods | Average | Aotra | Gallbladder | Kidney (L) | Kidney (R) | Liver | Pancreas | Spleen | Stomach |
|---|---|---|---|---|---|---|---|---|---|
| VNet | 68.81 | 75.34 | 51.87 | 77.10 | 80.75 | 87.84 | 40.04 | 80.56 | 56.98 |
| DARR | 69.77 | 74.74 | 53.77 | 72.31 | 73.24 | 94.08 | 54.18 | 89.90 | 45.96 |
| R50 U-Net | 74.68 | 87.74 | 63.66 | 80.60 | 78.19 | 93.74 | 56.90 | 85.87 | 74.16 |
| U-Net | 76.85 | 89.07 | 69.72 | 77.77 | 68.60 | 93.43 | 53.98 | 86.67 | 75.58 |
| R50 TransUNet | 75.57 | 55.92 | 63.91 | 79.20 | 72.71 | 93.56 | 49.37 | 87.19 | 74.95 |
| Att-UNet | 77.77 | 89.55 | 68.88 | 77.98 | 71.11 | 93.57 | 58.04 | 87.30 | 75.75 |
| VIT None | 61.50 | 44.38 | 39.59 | 67.46 | 62.94 | 89.21 | 43.14 | 75.45 | 68.78 |
| VIT CUP | 67.86 | 70.19 | 45.10 | 74.70 | 67.40 | 91.32 | 42.00 | 81.75 | 70.44 |
| R50 VIT CUP | 71.29 | 73.73 | 55.13 | 75.80 | 72.20 | 91.51 | 45.99 | 81.99 | 73.95 |
| R50-Deeplabv3+ | 75.73 | 86.18 | 60.42 | 81.18 | 75.27 | 92.86 | 51.06 | 88.69 | 70.19 |
| DualNorm-UNet | 80.37 | 86.52 | 55.51 | 88.64 | 86.29 | 95.64 | 55.91 | 94.62 | 79.80 |
| CGNET | 75.08 | 83.48 | 65.32 | 77.91 | 72.04 | 91.92 | 57.37 | 85.47 | 67.15 |
| ContextNet | 71.17 | 79.92 | 51.17 | 77.58 | 72.04 | 91.74 | 43.78 | 86.65 | 66.51 |
| DABNet | 74.91 | 85.01 | 56.89 | 77.84 | 72.45 | 93.05 | 54.39 | 88.23 | 71.45 |
| EDANet | 75.43 | 84.35 | 62.31 | 76.16 | 71.65 | 93.20 | 53.19 | 85.47 | 77.12 |
| ENet | 77.63 | 85.13 | 64.91 | 81.10 | 77.26 | 93.37 | 57.83 | 87.03 | 74.41 |
| FPENet | 68.67 | 78.98 | 56.35 | 74.54 | 64.36 | 90.86 | 40.60 | 78.30 | 65.35 |
| FSSNet | 74.59 | 82.87 | 64.06 | 78.03 | 69.63 | 92.52 | 53.10 | 85.65 | 70.86 |
| SQNet | 73.76 | 83.55 | 61.17 | 76.87 | 69.40 | 91.53 | 56.55 | 85.82 | 65.24 |
| FastSCNN | 70.53 | 77.79 | 55.96 | 73.61 | 67.38 | 91.68 | 44.54 | 84.51 | 68.76 |
| TransUNet | 77.48 | 87.23 | 63.16 | 81.87 | 77.02 | 94.08 | 55.86 | 85.08 | 75.62 |

| | | | | | | | | |
|---|---|---|---|---|---|---|---|---|
| SwinUNet | 79.13 | 85.47 | 66.53 | 83.28 | 79.61 | 94.29 | 56.58 | 90.66 | 76.60 |
| TransClaw U-Net | 78.09 | 85.87 | 61.38 | 84.83 | 79.36 | 94.28 | 57.65 | 87.74 | 73.55 |
| LeVit-UNet-384s | 78.53 | 87.33 | 62.23 | 84.61 | 80.25 | 93.11 | 59.07 | 88.86 | 72.76 |
| WAD | 80.30 | 87.73 | 69.93 | 83.95 | 79.78 | 93.95 | 61.02 | 88.86 | 77.16 |
| nnUNet | 86.99 | **93.01** | 71.77 | 85.57 | 88.18 | **97.23** | **83.01** | 91.86 | 85.25 |
| MS-Twins | **88.09** | 92.38 | **73.13** | **86.55** | **88.93** | 97.02 | 82.79 | **92.87** | **91.05** |

### 4.3 Experiments on ACDC

Table 2 shows experimental consequences on ACDC, in which the whole performance of the transformers-based baseline is better than that of the convent-based baseline. The root cause is that the pictures from ACDC have much fewer slices on the Z-axis (i.e., in Fig. 1, the space on the Z-axis is very large), which is an example where transformers have more advantages because they are conducted to process 2D input with less exchange on the Z-axis. As can be observed in Table 2, the optimal transformers model is Levit-UNET-384s, and its mean dice is a bit higher than SwinUNet but much higher than dual Attn on account of convnet. By comparison, MS-Twins on average exceeded Levit-UNET-384s by almost 1.5%, once again showing its superiorities over transformers-based baselines.

Table 2: Experiments on ACDC (dice mark in %, Best consequences are bolded.)

| Methods | Average | RV | Myo | LV |
|---|---|---|---|---|
| R50-U-Net | 87.55 | 87.10 | 80.63 | 94.92 |
| R50-Attn UNet | 86.75 | 87.58 | 79.20 | 93.47 |
| VIT-CUP | 81.45 | 81.46 | 70.71 | 92.18 |
| R50-VIT-CUP | 87.57 | 86.07 | 81.88 | 94.75 |
| TransUNet | 89.71 | 88.86 | 84.54 | 95.73 |
| SwinUNet | 90.00 | 88.55 | 85.62 | 95.83 |
| LeViT-UNet-384s | 90.32 | 89.55 | 87.64 | 93.76 |
| nnUNet | 91.59 | **90.25** | 89.10 | 95.41 |
| MS-Twins | **91.83** | 89.88 | **89.47** | **96.14** |

### 4.4 Ablation study

In this part, this study introduces the significance of the multi-scale feature iterative fusion module and cascading feature pyramid. In addition, this work also studies the effect of pretraining based on natural images on coding.

**Multi-Scale Feature Iterative Fusion block.** To study the influence of the multi-scale feature iterative fusion module, this study also deletes the sub-module for the experiment. As can be observed in Table 3, the result of iterative fusion modules with multi-scale features exceeds that of those without such modules by an average of nearly 3%.

Table 3: Research on multi-scale feature iterative fusion module

| | Average | Aotra | Gallbladder | Kidney(L) | Kidney(R) | Liver | Pancreas | Spleen | Stomach |
|---|---|---|---|---|---|---|---|---|---|
| No MS-FIF | 86.78 | 90.32 | 72.01 | 86.33 | 87.97 | 96.89 | 82.04 | 90.97 | 87.71 |
| MS-Twins | **88.09** | **92.38** | **73.13** | **86.55** | **88.93** | **97.02** | **82.79** | **92.87** | **91.05** |

**Influences of Cascade feature pyramid.** In Table 4, the consequences of replacing cascading feature pyramids in MS-Twins with convolution subsampling blocks are listed. As can be seen from Table 4, the average improvement of convolution falling sampling block is more than 3% compared with adjacent cascade, indicating that the application of cascade feature extraction is more conducive to the construction of hierarchical object concepts of different scales.

Table 4: A study of the cascading feature pyramid

|  | Average | Aotra | Gallbladder | Kidney(L) | Kidney(R) | Liver | Pancreas | Spleen | Stomach |
|---|---|---|---|---|---|---|---|---|---|
| Downsampling | 85.77 | 88.69 | 69.67 | 86.25 | 87.87 | 96.73 | 80.62 | 86.83 | 89.68 |
| MS-Twins | **88.09** | **92.38** | **73.13** | **86.55** | **88.93** | **97.02** | **82.79** | **92.87** | **91.05** |

**Influences of using pre-drilled models on natural graphics.** In Table 5, the application of pre-drilled weights on natural pictures is critical, which the removal of pre-trained weights reduces the whole partition performance by more than 3%. The root cause was that Synapse did not have sufficient tag scanning to thoroughly exploit the capacity of MS-Twins.

Table 5: Merits of using pre-drilled weights on natural pictures

|  | Average | Aotra | Gallbladder | Kidnery(L) | Kidnery(R) | Liver | Pancreas | Spleen | Stomach |
|---|---|---|---|---|---|---|---|---|---|
| No pre-training | 85.79 | 90.05 | 70.14 | 85.66 | 87.38 | 96.43 | 80.67 | 86.40 | 89.59 |
| MS-Twins | **88.09** | **92.38** | **73.13** | **86.55** | **88.93** | **97.02** | **82.79** | **92.87** | **91.05** |

## 4.5 Visualization

In Figure 2, the segmentation results of nnUNet and MS-twins in some samples are compared. In Synapse, the MS-Twins seem to have quite a distinct advantage in the stomach, while nnUNet often fails to generate a complete descriptive mask. At the same time, MS-twins could reduce the false positive prediction of spleen compared to nnUNet, which was also in accordance with the performance covered in Table 1.

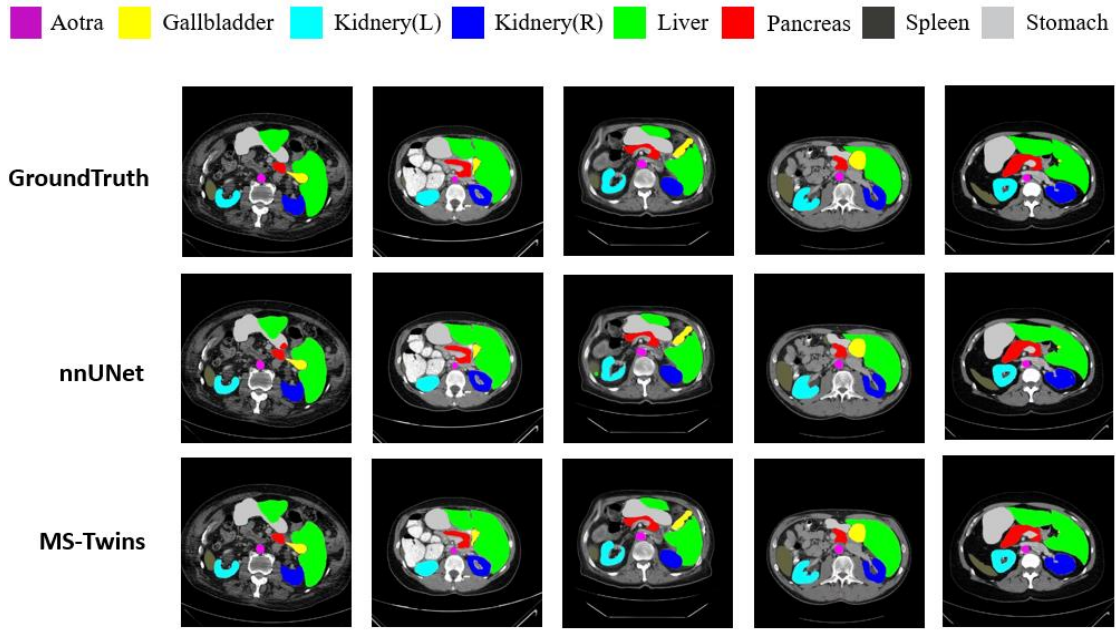

Fig. 2: Segmentation consequences of some hard samples on Synapse.

## 5 Conclusion

In the research, a novel medicine graphic partition network MS-Twins can be proposed. MS-Twins is built on the mixed backbone of convolution and self-attention. Convolution is conducive to encode accurate spatial data into high-resolution low-level characters and construct the concept of hierarchical objects on multiple scales. From another perspective, self-attention in the Transformer block combines long-term reply on convolution representations to catch the universal context. On account of this hybrid framework, MS-Twins has made great progress compared with the previous Transformers-based segmentation methods. Even compared to nnUNet, the best performing segmented network today, MS-Twins also gives coherence but noticeable progress. In future, this study believe that MS-Twins can attract more attention in the field of medical imaging and strive to develop a more effective segmentation model.

## 6 Acknowledgement

This study thanks NVIDIA Corporation for the GPU donation.